\documentclass[10pt, letterpaper]{article}

\usepackage[utf8]{inputenc}
\usepackage{amsmath, amssymb, amsfonts}
\usepackage{graphicx}
\usepackage{geometry}
\usepackage{cite}
\usepackage{hyperref}
\usepackage{cleveref}
\usepackage{svg}
\usepackage{enumitem}

\title{\textbf{Optimizing Network Topology Efficiency: \\ A Resource-Centric Analysis of Non-Blocking Architectures}}
\author{
  Jia Xu Wei \\
  \textit{University of California, Davis}
  \and
  Wei Wei \\
  \textit{Pleasanton, California}
}
\date{\today}

\begin{document}

\maketitle

\begin{abstract}
In modern network design, ``efficiency'' is often conflated with raw performance metrics like latency or aggregate throughput. This paper proposes a resource-centric definition of efficiency, isolating the hardware cost required to maintain a non-blocking throughput constraint. By modeling network cost as a function of the Traffic Multiplier (Hop Count) and Router Complexity (Radix), we demonstrate that the optimal topology is determined by the technological ratio between link interface costs ($\alpha$), crossbar switching costs ($\beta$), and the network concentration ratio. We conclude that while high-radix direct networks optimize efficiency at small to medium scales, indirect networks (e.g., Fat Trees) are required to cap router complexity at massive scales. Furthermore, we posit that redundancy is most efficiently handled via parallel network instances (e.g., multi-plane Star networks) rather than intrinsic topological path diversity.
\end{abstract}

\section{Introduction}
The evaluation of network topology for Data Centers and Network-on-Chip (NoC) architectures typically involves multi-dimensional trade-offs between latency, throughput, redundancy, and power. However, broadly comparing topologies with different blocking characteristics often leads to ambiguous results.

To isolate the \textit{efficiency} of the topology itself, this paper enforces a strict \textbf{Non-Blocking Constraint}: any candidate topology must provide sufficient bandwidth to sustain the full injection rate of all hosts under uniform traffic. Under this constraint, efficiency is defined strictly as the minimization of network resources (buffer area, crossbar logic, and link bandwidth) required to achieve this state. In this way, all network topologies are compared under the same effective capacity. 

\begin{figure}
    \centering
    \includegraphics[width=1\linewidth]{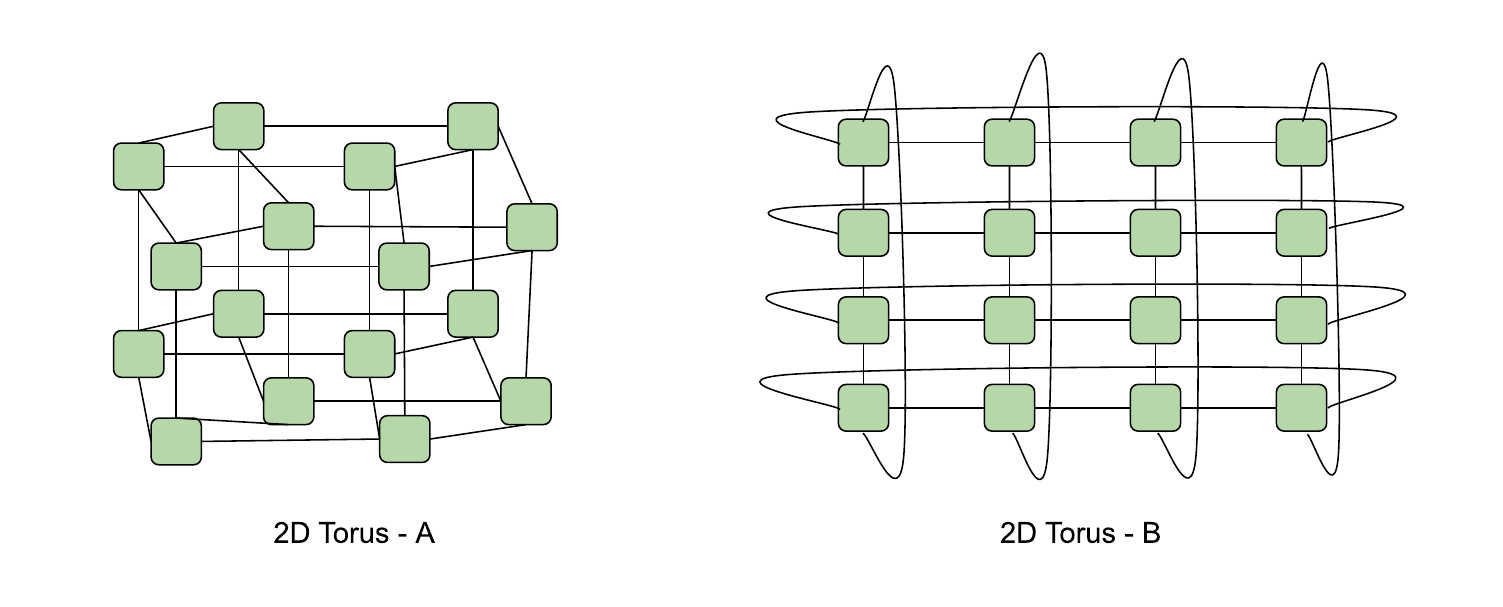}
    \caption{2D Torus}
    \label{fig:2d_torus}
\end{figure}

We limit our scope to the logical topology and node architecture. While physical layout and wire delay are significant implementation factors, they are treated here as uniform constraints applicable to all topologies \cite{dally_book}. This distinction is illustrated by the 2D torus configurations in \cref{fig:2d_torus}. Although Torus-A (folded) eliminates the long wraparound cables present in Torus-B (standard), they share the same graph connectivity and are therefore treated as the same logical network in this analysis.

\section{Theoretical Framework}
We define the network as a graph $G(N, L)$ serving a set of $M$ hosts.
\begin{itemize}
    \item \textbf{Hosts ($M$):} Terminals (Servers, Cores, or Processing Units) acting as traffic sources and sinks.
    \item \textbf{Nodes ($N$):} Switching elements (Routers, Switches, or Network Components) acting as traffic forwarding entities.
    \item \textbf{Links ($L$):} Physical interconnections between nodes.
\end{itemize}

To simplify the analysis, we assume a uniform random traffic distribution. While practical workloads may exhibit non-uniform or localized patterns, uniform traffic serves as the standard baseline for dimensioning global bandwidth in non-blocking architectures. In this regime, the primary determinant of resource consumption is the \textbf{Traffic Multiplication Factor}, defined as the number of times a single byte must be switched (forwarded) to reach its destination.

\section{The Resource-Efficiency Model}
To quantify efficiency, we derive a unified cost function ($Cost_{host}$) that represents the hardware burden per unit of host bandwidth.

\subsection{The Traffic Multiplier ($H$)}
In a packet-switched network, a data packet consumes bandwidth on every link it traverses. Therefore, the hop count functions as a \textbf{traffic multiplier}. The total volume of traffic ($V_{total}$) generated by the system is not merely the sum of the host injection bandwidth, but the injection bandwidth multiplied by the number of hops.

We define $H$ as the representative hop count, determined by the specific design constraints:
\begin{itemize}
    \item \textbf{Worst-Case Design:} Used to guarantee zero congestion in worst-case scenarios. This is applicable to most modern high-performance network designs, such as those in AI infrastructure.
    \begin{equation}
        H = D
    \end{equation}
where $D$ is the diameter of the network.
    \item \textbf{Average-Case Design:} Used for typical uniform traffic loads, often applied in traditional probabilistic network analysis.
    \begin{equation}
    H = L_{avg}
\end{equation}
where $L_{avg}$ is the average path length.
\end{itemize}
In practice, $H$ may also be determined by specific traffic models if required by unique network design requirements.

\subsection{Required Link Bandwidth ($BW_{req}$)}
To ensure the network is non-blocking, the physical links connecting the nodes must possess sufficient bandwidth to handle the multiplied traffic volume. Let $M$ be the total number of hosts. Let $L_{host}$ be the number of network links between nodes, normalized per host.
\begin{equation}
    L_{host} = \frac{2 \cdot L}{M}
\end{equation}
Where $L$ is the total number of inter-node links in the network (links between nodes and hosts are excluded). The factor of 2 accounts for the full-duplex operation.

Let $B_{host}$ be the normalized injection bandwidth per host (throughput unit). The aggregate bandwidth demand generated by the system is $M \times B_{host} \times H$. This load is distributed across the network's available links. The required bandwidth per network link ($BW_{req}$) is, therefore, derived as:
\begin{equation}
    BW_{req} =
    \begin{cases}
        \frac{H}{L_{host}} \cdot B_{host}, & \text{if } H \ge L_{host} \\
        \\
        B_{host}, & \text{if } H < L_{host} \text{ as in over provisioned network}        
    \end{cases}    
\end{equation}
\textbf{Assumption:} This model assumes uniformly distributed traffic. In practice, this assumption is readily fulfilled and holds true for the dimensioning of modern non-blocking architectures.

This derivation implies that topologies with large diameters (e.g., Ring, Mesh) require significantly wider or faster links to prevent saturation compared to low-diameter topologies \cite{kim_butterfly}.

\subsection{Router Complexity Cost ($C_{router}$)}
The cost of the switching node (router) is the dominant non-linear variable. Following the analysis by Kim and Dally \cite{kim_isca05}, we model the router cost as a function of its radix ($k$):
\begin{equation}
    C_{router}(k) = \alpha \cdot k + \beta \cdot k^2
\end{equation}

\begin{itemize}
    \item $\alpha \cdot k$ \textbf{(Port Cost):} Linearly scaling components such as SerDes, link drivers, and flow-control buffers.
    \item $\beta \cdot k^2$ \textbf{(Switching Cost):} Quadratically scaling components, primarily the crossbar matrix and arbitration logic \cite{kim_isca05}. In standard matrix crossbars, the number of crosspoints increases as $k^2$, dominating the area and the power budget at high radices.
\end{itemize}

Since our objective is to compare the relative efficiency of different network topologies, we normalize the cost function. $C_{router}$ can be expressed as:
\begin{equation}
    C_{router}(k) = k + \frac{\beta}{\alpha} \cdot k^2
\end{equation}
In this paper, we will adopt this normalized function.

\textit{Note:} The ratio $\alpha/\beta$ represents the ``technology crossover point'' where the switch core becomes more expensive than the aggregate cost of link interfaces and forwarding engines.

\subsection{The Unified Efficiency Function ($Cost_{host}$)}
We define the system efficiency metric as the **Resource Cost per Host Bandwidth** needed to sustain non-blocking performance. This combines the link bandwidth requirement (driven by the topology) with the router cost (driven by the degree of the node).

Let $N$ be the number of network nodes (routers). The ratio $\frac{N}{M}$ represents the number of switching nodes required per host (Network Concentration).

Combining these factors, the total cost of resources per host bandwidth ($B_{host}$) is as follows:
\begin{equation}
    Cost_{host} = 
    \begin{cases}
      \frac{H}{L_{host}} \times (k + \frac{\beta}{\alpha} k^2) \times \left( \frac{N}{M} \right), & \text{if } H \ge L_{host} \\
      \\
      (k + \frac{\beta}{\alpha} k^2) \times \left( \frac{N}{M} \right), & \text{if } H < L_{host} \text{ as in over provisioned network}
    \end{cases}    
\end{equation}
where:
\begin{itemize}
    \item $H$: Traffic Multiplier (Hop Count)
    \item $L_{host}$: Network links per host
    \item $(k + \frac{\beta}{\alpha} k^2)$: Router Complexity
    \item $N/M$: Network Concentration (Nodes per Host)
\end{itemize}
Efficiency is maximized by minimizing this value.

\subsection{Analysis of the Trade-off}
This equation reveals the fundamental tension in topology design:

\begin{enumerate}
    \item \textbf{To reduce $BW_{req}$:} We must reduce $H$ (hops). This typically requires increasing the connectivity (radix $k$) of the routers (e.g., moving from a Torus to a Hypercube).
    \item \textbf{To reduce $C_{router}$:} We must keep $k$ low to avoid the quadratic penalty $\frac{\beta}{\alpha} k^2$. This typically forces a higher hop count $H$ (e.g., moving from a Hypercube back to a Ring/Torus).
    \item \textbf{Network Concentration Tradeoff:} In indirect networks, the ratio $\frac{N}{M}$ acts as a distinct variable to manage the tradeoff for the radix $k$. This allows the topology to be optimized for a specific technology ratio $\alpha/\beta$.
\end{enumerate}

The optimal topology is found by minimizing $Cost_{host}$, achieved by balancing $H$ against $k^2$, utilizing $\frac{N}{M}$ in indirect networks as a scaling lever.

\section{Comparative Topology Analysis}
In modern router microarchitectures, the Linear Cost ($\alpha$)—dominated by SerDes (Serializers/Deserializers), signal drivers, and per-port buffering (SRAM)—often dwarfs the Quadratic Cost ($\beta$) of the crossbar logic, especially at low-to-medium radices. Although the digital logic of a crossbar scales quadratically, these gates are relatively inexpensive compared to the power- and area-hungry analog/mixed-signal components and complex logic required at the port interfaces.

To evaluate the efficiency of standard topologies, we apply the Unified Efficiency Function derived in Section III. We specifically examine the ``Low-Beta Regime'' ($\beta \ll \alpha$), where the cost of adding a port (linear penalty) initially outweighs the complexity of switching between them (quadratic penalty).

For this comparison, we assume a Direct Network where the number of Nodes ($N$) is equal to the number of Hosts ($M$) (Ratio = 1) to isolate topological differences.

\subsection{The Comparison Classes}
We analyze three classes of topology that represent the spectrum of Radix ($k$) vs. Hop Count ($H$).

\subsubsection{Low Radix / High Diameter (e.g., 2D Torus)}

\textbf{Configuration:} Fixed low radix ($k = 5$; 4 inter-node links, 1 host link), Fixed links per host $L_{host}=4$ and High diameter ($H = \sqrt{N}$ if $\sqrt{N}$ is even; $H = \sqrt{N} - 1$ if $\sqrt{N}$ is odd).
\\
\textbf{Analysis:} Since $k$ is small and constant, the router cost is minimized and dominated by the linear term $k$. The quadratic penalty $\frac{\beta}{\alpha} k^2$ is negligible.

However, although the cost of the router is minimal ($5 + 25\frac{\beta}{\alpha}$), the Traffic Multiplier $H$ grows rapidly on scale. To remain non-blocking, a Torus requires massive link bandwidth over-provisioning.\begin{equation*}
    Cost_{2dTorus}=
    \begin{cases}
        (5+25\frac{\beta}{\alpha}) \cdot \frac{\sqrt{N}}{4}, & \text{if } \sqrt{N} \text{ is even and } \ge 4 \\
        \\
        (5+25\frac{\beta}{\alpha}) \cdot \frac{\sqrt{N} - 1}{4}, & \text{if } \sqrt{N} \text{ is odd and } \ge 5 \\
        \\
        5+25\frac{\beta}{\alpha}, & if \sqrt{N} < 4
    \end{cases}
\end{equation*}
\textbf{Verdict:} \textit{Inefficient} for large $N$ due to the high traffic multiplication factor. The cost scales at $O(\sqrt{N})$ with a factor of $\frac{1}{4}$.

\subsubsection{Medium Radix / Logarithmic Diameter (e.g., Hypercube)}
\textbf{Configuration:} Radix scales logarithmically ($k=log_2N+1$; $log_2N$ inter-node links, 1 host link), Diameter scales logarithmically ($H = log_2N$), and $L_{host} = H$. 
\\
\textbf{Analysis:} Both $H$ and $k$ grow, but at a moderate logarithmic rate.
\begin{equation*}
    Cost_{Hypercube} = (log_2N+1) + \frac{\beta}{\alpha}(log_2N+1)^2
\end{equation*}
\\
\textbf{Verdict:} \textit{Scalable Efficiency}. The cost scales roughly at $O(log_2N)$ in the low-$\frac{\beta}{\alpha}$ regime for medium values of $N$.

\subsubsection{High Radix / Low Diameter (e.g., Flattened Butterfly)}
\textbf{Configuration:} Radix scales rapidly ($k = 2\sqrt{N}-1$), Fixed low diameter ($H = 2$), Network is over-provisioned ($H \le L_{host}$).
\\
\textbf{Analysis:} Using the ``Low-Beta'' regime, these topologies trade a linear increase in port cost for a reduction in hops \cite{kim_butterfly}. Although $H$ is small and fixed, the radix grows rapidly with scale, resulting in significant over-provisioning.
\begin{equation*}
    Cost_{FlattenedButterfly} = (2\sqrt{N}-1) + \frac{\beta}{\alpha}(2\sqrt{N}-1)^2
\end{equation*}
The cost grows at a rate approximately 8 times that of the 2D Torus case. Analysis indicates that it is cost-effective only at very small scales (e.g., $N \le 9$). 
\\
\textbf{Verdict:} \textit{Lowest Latency}, but at a high router cost even when $k$ remains below the technology limit. The cost scales roughly at $O(\sqrt{N})$ with a factor of 2.

\subsection{The Optimization Balance: Hop Count vs. Radix}
Optimizing the topology requires a careful balance between Hop Count ($H$) and Radix ($k$). Minimizing only one variable fails to produce an optimal result.
\begin{enumerate}
    \item \textbf{Inefficiency of Torus:} Although the Torus utilizes the ``cheapest'' routers, the Traffic Multiplier ($H \propto \sqrt{N}$) forces the system to provision massive amounts of extra bandwidth. In the cost function, the growth of the $\sqrt{N}$ term far exceeds the savings provided by the $\frac{\beta}{\alpha}$ term.
    \item \textbf{Inefficiency of Flattened Butterfly:} Although the Flattened Butterfly achieves the ``lowest'' Hop Count, the radix ($k \propto \sqrt{N}$) forces the system to provision high-radix routers with excessive bandwidth for over-provisioning. Similarly to Torus, the $\sqrt{N}$ growth term in the cost function outweighs the benefits of the $\frac{\beta}{\alpha}$ term.
    \item \textbf{The Hypercube Advantage:} Since $\frac{\beta}{\alpha}$ is small, the penalty for increasing the radix from 5 (Torus) to, for instance, 10 or 12 (Hypercube size) is largely logarithmic. We pay a small logarithmic price in router cost to gain a linear reduction in hop count (and thus the required link bandwidth).
\end{enumerate}

\section{Scale Dependence and Network Concentration}
No single topology maintains efficiency on all scales. The variable $N/M$ (Network Concentration) plays a critical role in decoupling the complexity of the router from the aggregate size of the system.

\subsection{The Limits of Direct Networks ($N = M$)}
In direct networks (e.g., Torus, Hypercube, Flattened Butterfly), the node count is tightly coupled to the host count. To maintain a low hop count $H$ as $M$ increases, the radix $k$ must increase rapidly, as demonstrated by the Flattened Butterfly topology. Eventually, $k$ becomes large enough that the quadratic term $\frac{\beta}{\alpha} k^2$ dominates, causing the hardware implementation cost to become prohibitive \cite{kim_isca05}.

\subsection{Indirect Networks ($N \neq M$)}
Indirect topologies, such as Fat Trees \cite{leiserson_fat_tree}, introduce switching-only nodes ($N \neq M$).  This increases the third term of our cost function ($N/M$) but allows us to limit the second term ($C_{router}$) by fixing $k$ at a constant optimal value (e.g., $k=64$) while maintaining a low $H$.

In this topology, the hop count is as $H = L_{host} = 2 \times (Level-1)$, where $Level$ represents the number of network layers.

Let $l=Level$. In a standard configuration, each router allocates half of its ports to the lower tier and half to the upper tier. Specifically, $k/2$ links connect to hosts or routers of the lower-layer, and $k/2$ links connect to routers of the upper-layer. The exception is the root (top) layer, where all links connect downward to lower-layer routers or hosts. This results in a one-layer setup that effectively functions as a Star topology.
\begin{equation*}
\begin{split}
    & 2 \cdot (\frac{k}{2})^l = M \\
    \Rightarrow & (\frac{k}{2})^l = \frac{M}{2} \\
    \Rightarrow & l \cdot log_2\frac{k}{2} = log_2\frac{M}{2} \\
    \Rightarrow & l=\frac{log_2M-1}{log_2k-1} \\
    \\
    N & = \frac{M}{k/2}(l-\frac{1}{2}) \\
    & = \frac{2M}{k} \left(\frac{log_2M-1}{log_2k-1} - \frac{1}{2} \right) \\
    \\
    H & = 2 \cdot (l-1) \\
    & = 2 \cdot \left(\frac{log_2M-1}{log_2k-1} - 1 \right)
\end{split}
\end{equation*}

Combining these factors, the Cost of the Fat Tree is derived as follows:
\begin{equation}
\begin{split}
    Cost_{FatTree} & = \frac{H}{L_{host}} \times C_{fixed}(k_{opt}) \times \left( \frac{N}{M} \right) \\
    & = C_{fixed}(k_{opt}) \times \left( \frac{N}{M} \right) \\
    & = (k + \frac{\beta}{\alpha}k^2) \times \frac{2}{k} \left(\frac{log_2M-1}{log_2k-1} - \frac{1}{2} \right) \\
    & = 2(1 + \frac{\beta}{\alpha}k)\left(\frac{log_2M-1}{log_2k-1} - \frac{1}{2} \right)
\end{split}
\end{equation}

This structure allows the network to scale indefinitely without any single router hitting the ``Quadratic Wall'' of complexity.

In the degenerate case of a one-layer setup (Star topology), $H = L_{host} = 0$. Without loss of generality (WLOG), we define $\frac{H}{L_{host}} = 1$ in this case. As $M = k$, the Cost simplifies to:
\begin{equation}
\begin{split}
    Cost_{Star} & = 2(1 + \frac{\beta}{\alpha}k)\left(\frac{log_2M-1}{log_2k-1} - \frac{1}{2} \right) \\
    & =2 \cdot (1 + \frac{\beta}{\alpha}k) \cdot \frac{1}{2} \\
    & = 1 + \frac{\beta}{\alpha}k
\end{split}
\end{equation}

\subsection{The Impact of $\beta \ll \alpha$ (The ``SerDes Dominated'' Reality)}
When we apply the insight that the crossbar logic ($\beta$) is significantly cheaper than the port interface ($\alpha$), the efficiency landscape shifts dramatically in favor of High Radix topologies over Low Radix ones.
\begin{enumerate}
    \item \textbf{Stable Router Cost Factor:} Given a small ratio $\frac{\beta}{\alpha}$, the term $\frac{\beta}{\alpha}k$ remains less than 1 even with a relatively high radix $k$. This makes the term ($1+\frac{\beta}{\alpha}k$) largely consistent as $k$ increases.
    \item \textbf{Reduction in Network Concentration ($\frac{N}{M}$):} The ratio $\frac{N}{M}$ decreases logarithmically as $k$ grows. Consequently, the total cost decreases in tandem with the hop count $H$, leading to improved performance and unified cost.
\end{enumerate}

\subsection{Synthesis: Why "High-Radix" Wins}
This analysis mathematically confirms the industry trend towards high-radix routers (e.g., Radix-64 or Radix-128 switches in modern data centers).

Because $\frac{\beta}{\alpha}$ is small (driven by Moore's Law making logic dense and cheap), it is economically viable to push $k$ quite high—well into the 30--60 range or beyond—before the $k^2$ penalty becomes prohibitive. This capability allows for the construction of networks with very low diameters ($H=2$ or $H=3$) that are significantly more resource-efficient than low-radix alternatives while achieving higher performance at the same time. This approach effectively utilizes ``cheap logic'' to conserve ``expensive bandwidth.'' For example, utilizing Radix-512 switches, a 2-layered Fat Tree network could theoretically connect 131,072 hosts without blocking. 

\section{Conclusion}
This paper has re-evaluated network topology efficiency by shifting the focus from raw performance benchmarks to a resource-centric cost analysis under a fixed non-blocking constraint. By normalizing all topologies to a uniform throughput standard, we isolated the underlying resource consumption required to support that performance.

This resource-centric analysis demonstrates that network efficiency is a balancing act between \textbf{Traffic Multiplier ($H$)} and \textbf{Switching Penalty ($k^2$)}.

\begin{enumerate}
    \item \textbf{Metric Validity:} Standard metrics like Bisection Bandwidth are insufficient for efficiency analysis unless coupled with the router complexity cost ($\beta k^2$) required to achieve them.
    \item \textbf{High-Radix Dominance:} In modern technology nodes, the logic penalty for switching is low ($\beta \ll \alpha$). Consequently, topologies that utilize high-radix routers to minimize hop counts are mathematically superior to low-radix designs \cite{kim_isca05, kim_butterfly}.
    \item \textbf{The Scaling Threshold:} For massive scales, Direct Networks become inefficient due to excessive radix requirements. The efficiency in this regime dictates the use of \textbf{Indirect Networks} \cite{leiserson_fat_tree}, which optimize the total cost by increasing the number of nodes ($N$) to maintain optimal router complexity ($k$).
    \item \textbf{Redundancy via Parallelism:} The necessity for intrinsic topological redundancy (internal path diversity) is diminished by modern host architectures. As hosts equipped with SmartNICs or native inter-chip links can connect to multiple disjoint networks simultaneously (e.g., parallel Star planes), system-level reliability is increasingly achieved through network parallelism rather than topological complexity. This paradigm shift justifies the exclusion of redundancy overhead from our primary efficiency function.
\end{enumerate}

In conclusion, the design of an efficient network is not about selecting a specific shape (Torus vs. Tree), but about solving the resource equation. The optimal design methodology is to first determine the maximum cost-effective router radix $k_{opt}$ based on the technology constants $\alpha$, $\beta$, and the number of hosts $M$, and then employ an indirect structure to scale that optimal building block to the desired number of hosts.


\end{document}